\documentclass[11pt]{article}
\usepackage{amsmath,amsfonts}
\usepackage{graphicx}
\def\sloppy{\tolerance=100000\hfuzz=\maxdimen\vfuzz=\maxdimen}
\def \be {\begin{equation}}
\def \ee  {\end{equation}}
\def \bear {\begin{eqnarray}}
\def \eear {\end{eqnarray}}
\def\sqr#1#2{{\vcenter{\vbox{\hrule height.#2pt
\hbox{\vrule width.#2pt height#1pt \kern#1pt
\vrule width.#2pt}\hrule height.#2pt}}}}

\def\la {{\langle}}
\def\ra {{\rangle}}

\def\Tr {{\rm Tr}}
\def \tr {{\rm tr}}

\def\del {\partial}

\def\half{\textstyle{1\over 2}}

\begin{document}
\fontfamily{cmr}\fontsize{11pt}{14pt}\selectfont
\def \CMP {{Commun. Math. Phys.}}
\def \PRL {{Phys. Rev. Lett.}}
\def \PL {{Phys. Lett.}}
\def \NPBProc {{Nucl. Phys. B (Proc. Suppl.)}}
\def \NP {{Nucl. Phys.}}
\def \RMP {{Rev. Mod. Phys.}}
\def \JGP {{J. Geom. Phys.}}
\def \CQG {{Class. Quant. Grav.}}
\def \MPL {{Mod. Phys. Lett.}}
\def \IJMP {{ Int. J. Mod. Phys.}}
\def \JHEP {{JHEP}}
\def \PR {{Phys. Rev.}}
\def \JMP {{J. Math. Phys.}}
\def \GRG{{Gen. Rel. Grav.}}
\begin{titlepage}
\null\vspace{-62pt} \pagestyle{empty}
\begin{center}
\rightline{CCNY-HEP-11/5}
\rightline{September 2011}
\vspace{1truein} {\Large\bfseries
Fuzzy Spaces and New Random Matrix Ensembles}\\
\vspace{6pt}
\vskip .1in
{\Large \bfseries  ~}\\
\vskip .1in
{\Large\bfseries ~}\\
{\large V.P. Nair, A.P. Polychronakos and J. Tekel}\\
\vskip .2in
{\itshape Physics Department\\
City College of the CUNY\\
New York, NY 10031}\\
\vskip .1in
\begin{tabular}{r l}
E-mail:
&{\fontfamily{cmtt}\fontsize{11pt}{15pt}\selectfont vpn@sci.ccny.cuny.edu}\\
&{\fontfamily{cmtt}\fontsize{11pt}{15pt}\selectfont alexios@sci.ccny.cuny.edu}\\
&{\fontfamily{cmtt}\fontsize{11pt}{15pt}\selectfont tekel@sci.ccny.cuny.edu}
\end{tabular}

\fontfamily{cmr}\fontsize{11pt}{15pt}\selectfont
\vspace{.8in}
\centerline{\large\bf Abstract}
\end{center}
We analyze the expectation value of observables in a scalar theory on the fuzzy two sphere, represented as a
generalized hermitian matrix model. We calculate explicitly the form of the expectation values in the large-$N$ limit and demonstrate that, for any single kind of field (matrix), the  distribution of its eigenvalues is still a Wigner semicircle but with a renormalized
radius. For observables involving more than one type of matrix we obtain a new distribution corresponding to correlated
Wigner semicircles.
\end{titlepage}
\section{Introduction}

Matrix models have been of interest for a long time going back to Wigner's work defining the
classic Gaussian ensembles for matrices \cite{wigner}. 
The motivation for this was to understand the distribution of energy levels 
for heavy nuclei, for which the Hamiltonian has so many contributing interactions that it
my be taken to be a random $N\times N$ matrix. The distribution of eigenvalues in the large $N$ limit could then be compared to the distribution of energy levels for a large collection of different nuclei. Assuming a normal distribution for the elements of the matrix,
the classic semicircle law of Wigner emerged in this context.
Since then matrix models have emerged in many other contexts in physics: in modeling Riemann surfaces with a view to applications in string theory \cite{string}, in integrable systems, such as the Calogero model, as well as other condensed matter systems \cite{guhr}, in possibly elucidating the concept of chaos in quantum systems \cite{chaos}, etc.

Since matrices are a simple example of noncommuting variables, Wigner's matrix ensembles
also turn up in the noncommutative probability theory of Voiculescu \cite{voic}. In fact, the semicircle law plays a very important role in free (uncorrelated noncommutative) probability theory. Essentially it is to free probability theory what the normal distribution is to the probability theory of commuting variables, with a corresponding central limit theorem \cite{voic}.

An obvious question that arises is whether there are other matrix ensembles of importance in physics and mathematics 
which are also naturally defined.
Fuzzy spaces are almost a self-evident answer to this question.
Noncommutative (fuzzy) spaces, and field theories on such spaces,
have been an important topic of research for a long time now \cite{douglas}.
Such spaces can arise as brane solutions in certain contexts
in string theory and in the matrix version of $M$-theory \cite{taylor}.
Gauge theories on such spaces are interesting since they can describe fluctuations of the brane solutions and unify in a natural
way gauge and spatial degrees of freedom. This has generated interest more generally on field theories on fuzzy and noncommutative spaces.

Fuzzy spaces are noncommutative spaces that can be described by finite dimensional matrices, and, by now, there are many examples of such spaces.
When the dimension of the matrices becomes large, these spaces tend to corresponding commutative manifolds in terms of their geometry and the algebra of functions on such spaces.
From the physics point of view fuzzy spaces are important for many reasons: 1) They provide
a regularization that can preserve various symmetries, even avoiding the fermion doubling problem as compared to the standard
lattice regulator. 2) They do play a role as effective descriptions of certain condensed matter systems such as the quantum Hall effect. 3) Being a finite-mode approximation to fields preserving isometries, they have implications for quantum gravity.
4) Two dimensional YM theory naturally reduces to a (unitary) matrix model with time as a continuous parameter.

There are, of course, many diverse issues here, but if we take the simple case of a field theory on a fuzzy space, with the fields being $N \times N$ matrices, the Euclidean functional integral is a matrix ensemble. The Gaussian ensembles of Wigner correspond to mass terms for the fields.
The action also has kinetic energy terms, given, for a scalar field, by the matrix Laplacian.
The Laplacian essentially defines the geometry of the continuum manifold as we take the large $N$ limit and hence it is the key geometrical ingredient.
This clearly gives a natural class of matrix ensembles. So, unlike the Wigner distribution, we must seek a large $N$ limit which takes account of  the contribution of the Laplacian. We also have new observables in this case and the Laplacian will give some degree of correlation which is reflective of the emergent geometry at large $N$.

Not surprisingly, there have been attempts to understand the role of and generalize the Wigner distribution for fuzzy spaces \cite{{ocon}, {stein}}.
The authors of \cite{ocon}, in particular, start with the mass term as the leading term of the action and integrate out the angular modes which appear in the kinetic term, treating this term as a perturbation.
There is, however, no guarantee of continuously connecting this to the zero mass limit because of
possible phase transitions as a function of $N$ and the lack of summability of the perturbation series. Hence an approach treating all terms of the action on an equal footing with the possibility of different scaling limits is desirable.

The formulation of scalar field theory on the fuzzy sphere by
Steinacker, on the other hand, is closer to the spirit of our work.
Using momentum space integration, rather than matrix techniques, the
Wigner distribution for the eigenvalues of the matrix is recovered
\cite{stein}. The main difference with the present work is our focus
on observables containing also derivatives of the field (Laplacians
and their powers). These manifest the dependence of the results on the
angular variables of the matrix and introduce nontrivial correlations.
The calculation of mixed expectation values, with both powers of the
field and its derivatives, and the resulting correlated Wigner
distribution are our main results.

\section{Ensembles and distribution functions}

\subsection{The ensembles}

We start with the simplest case of a real scalar field on a fuzzy two-sphere.
The Cartesian coordinates are $N \times N$ matrices:
\be
X_\alpha = \frac{2r}{N} L_\alpha ~, ~~ [L_\alpha , L_\beta ] = i \epsilon_{\alpha \beta \gamma} L_\gamma ~,~~
\sum_\alpha X_\alpha^2 = \left( 1 - \frac{1}{N^2} \right) r^2
\ee
where $L_\alpha$, $\alpha = 1, 2, 3$ are $SU(2)$ generators (angular momentum matrices) in the $N$-dimensional representation,
$r$ represents the radius of the sphere and $\theta = 2 r^2/N$ the noncommutativity parameter.
Fields become general $N \times N$ matrices $M$. Derivatives (rotations) and the corresponding Laplacian are $L$-commutators
\be
{\cal L}_\alpha M = -i [L_\alpha , M] ~,~~~ \Delta M = - \frac{1}{r^2} \sum_\alpha [L_\alpha , [L_\alpha , M]]
\ee
while integration over the sphere becomes a matrix trace
\be
\int_{S^2} d^2 x \, \Phi = \frac{4\pi r^2}{N} \tr M
\ee
A real free scalar field on the two-sphere is represented by a hermitian $N \times N$ matrix $M$.
Upon rescaling $M$ to normalize the kinetic term, 
the Euclidean action is given by
\bear
S &=& - {1\over 2} \Tr \big( [L_\alpha, M]\, [L_\alpha, M] \big) ~+~ {\mu^2 \over 2} \Tr \big(M^2\big)\nonumber\\
&=&  ~{1\over 2} \Tr \big(M\, [L_\alpha, [L_\alpha, M]] \big) ~+~ {\mu^2 \over 2} \Tr \big(M^2\big)
\label{1}
\eear
We shall be interested in the large-$N$ limit of (\ref{1}). Classically, for `smooth' configurations,
$M$ is replaced by its symbol in this limit, which is now a real scalar field 
$\phi$ on the two-sphere. The adjoint action of $L_\alpha$ on $M$, i.e. $[L_\alpha , M]$, becomes the gradient of $\phi$ and the action, upon appropriate scaling, becomes the free field action on the two-sphere,
\be
S = {1\over 2} \int d\mu (S^2)\,  \left[ (\nabla \phi )^2 + \mu^2 \phi^2 \right]
\label{2}
\ee
Quantum mechanically, however, expectation values of observables depend on the noncommutativity parameter,
which acts as a regulator of infinities in the continuum, and the large-$N$ limit remains nontrivial.

For the usual Gaussian matrix ensembles for which we just have the mass term in
(\ref{1}), we have the symmetry $M \rightarrow U^\dagger M \, U$ for a unitary matrix 
$U$, which allows us to diagonalize $M$ as $ M = U^\dagger  M_{diag} U$.
Only the eigenvalues of $M$ are physically meaningful and their distribution is given in the large $N$ limit by the Wigner distribution function. The observables of interest are of the form
$\Tr M^k$.
In the case of (\ref{1}), it is still possible to scale variables and the parameter $\mu$ in such a way that the Laplacian of $M$,
which we denote by $B = [L_\alpha, [L_\alpha, M]]$, becomes irrelevant in the large $N$ limit and one recovers just the Wigner result.
However, we are interested in cases where the gradient term is not irrelevant, particularly for possible application to field theory on fuzzy spaces.
In fact, we will consider an action which is more general than the one given in
 (\ref{1}),
 \be
S=  {1\over 2} \Tr (M\, K\, M ) ~+~ {\mu^2 \over 2} \Tr (M^2)
\label{3}
\ee
where the kinetic operator $K$ can be more general than $(L^{adj}_\alpha)^2$. 

The observables involve arbitrary products of $M$, $B = K\, M$, $[L_\alpha, M]$, etc.
Rather than changing variables to the eigenvalues ($M_{diag}$) and the angular degrees of freedom ($U$), we will consider the moment generating functions for the observables and obtain the distributions functions which lead to them. 
For this, the form of the action itself will not be very important; in fact
(\ref{3}) is meant more as a guide or motivation.
All we need is that correlators can be evaluated by Wick contractions in terms of products of two-point functions. Rather than the action, we can specify the scaling behavior of two-point functions. We will examine possible scalings which give a finite large $N$ limit.

\subsection{Recursion rules and the generating function for correlators}

To facilitate the upcoming calculations,
we define a basis for $N\times N$ matrices $\{ T^{(l)}_A\}$, where
$A = 1, 2, \cdots , 2 l+1$ for each $l$ and $l = 0, 1, \cdots, N-1$.
These matrices $T^{(l)}_{A}$ transform
as the spin-$l$ representation of $SU(2)$. Explicitly, a possible choice for the elements of this base is
${\bf 1}$, $L_\alpha$, $L_\alpha L_\beta + L_\beta L_\alpha - (2/3) L^2 \delta_{\alpha\beta}$,
etc.
We take $T^{(l)}_A$ to be normalized as
\be
\Tr \,\left(T^{(l)}_A\, T^{(l')}_B\right) = \delta^{l l'} \, \delta_{AB}
\label{4}
\ee
Further, $[L_\alpha, [L_\alpha, T^{(l)}_A]] = l (l+1)\, T^{(l)}_A$. The matrix $M$ may be expanded in this basis as $M = \sum_{l,\, A}c_A^l \, T^{(l)}_A$. The action is diagonal in terms of this basis,
with the expectation value $\la c^l_A c^{l'}_B \ra = \delta_{AB} \,\delta^{l l'} \, G(l)$.
For the action (\ref{1}), $G(l) = [\mu^2 + l (l+1)]^{-1}$, but it can be taken to be a more general function of $l$ depending
on the form of the kinetic term operator $K$ in (\ref{3}). Using the fact that the matrices $T^{(l)}_A$ transform as the spin-$l$ representation of
$SU(2)$, we see that
\be
\sum_A \left(T^{(l)}_A \, T^{(l)}_A \right)_{ij} = {2 l +1 \over N}\, \delta_{ij}
\label{5}
\ee
This follows easily from carrying out an $SU(2)$ transformation on the states
corresponding to the matrix labels $i$, $j$.

First consider the propagator for $M$, or the two-point function $\la M_{ij} M_{kl} \ra$.
As a result of relation (\ref{5}), the two-point function (or propagator) for a matrix product of two matrices takes the form
\be
\la \, ( M \, M )_{ij} \ra = {1\over N} \sum_{l=0}^{N-1} (2 l +1) G(l)\, \delta_{ij} \equiv f\, \delta_{ij}
\label{6}
\ee
Similarly, we can define the propagators
\be
\la ( B\, B )_{ij} \ra = g\, \delta_{ij}, \hskip .3in \la ( M \, B )_{ij} \ra = \la ( B \, M )_{ij} \ra =  h \, \delta_{ij}
\label{7}
\ee
For $B = [L_\alpha , [L_\alpha , M]]$, we can explicitly obtain
\be
g = {1\over N} \sum_{l=0}^{N-1}  (2 l+1) l^2 (l+1)^2 G(l), \hskip .3in
h = {1\over N} \sum_{l=0}^{N-1}  (2 l+1) l (l+1) G(l)
\label{8}
\ee
Note that, for positive $G(l)$, an application of the Schwartz inequality to the above sums gives the relation
\be
f g \ge h^2
\label{7a}
\ee

An important result in the large $N$ limit is that, for any matrix product, the expectation value is proportional to the identity
matrix. To see this, consider $\la A_{ij} \ra$ where $A$ is a matrix of the form
\be
A = M^{m_1} \, B^{b_1} \, M^{m_2} \, B^{b_2} \cdots
\label{9}
\ee
In evaluating the expectation value of $A_{ij}$ by Wick contractions, the leading term in the large $N$ limit will involve only planar contractions; i.e., diagrams where the the propagators ``cross" are subleading in $1/N$.
(This can be shown by examining the scaling
of the matrix elements for $T^{(l)}_A$ that appear in such contractions and the corresponding factors of $N$ arising from
index summations.) As a result, in the leading term, there will be at least one case where two adjacent matrices are contracted.
Denoting $Q_1 Q_2$ the two adjacent matrices, and $\cal L$ and $\cal R$ the matrices to the left and right of $Q_1 Q_2$,
and using (\ref{6},\ref{7}), this contraction will give us a term of the form
\be
\la A_{ij} \ra \approx \la {\cal L}_{ia}\, {\cal R}_{bj}\ra \, \la (Q_1 Q_2)_{ab}\ra
\sim \la ( {\cal L} {\cal R})_{ij}\ra \, q
\label{10}
\ee
where $Q_1$, $Q_2$ stand for either $M$ or $B$ and $q$ is the appropriate function
among $f$, $g$, $h$.
This step reduces the correlator to one with lower number of 
matrices $M$ or $B$, with the two nearby matrices that were Wick contracted deleted.
 Iterating, we see that
\be
\la A_{ij} \ra = a\, \delta_{ij}
\label{11}
\ee
where $a = \la \tr A \ra /N$ is a scalar quantity.

\begin{figure} [h,t]                     
\begin{center}
\includegraphics[width=0.8\textwidth]{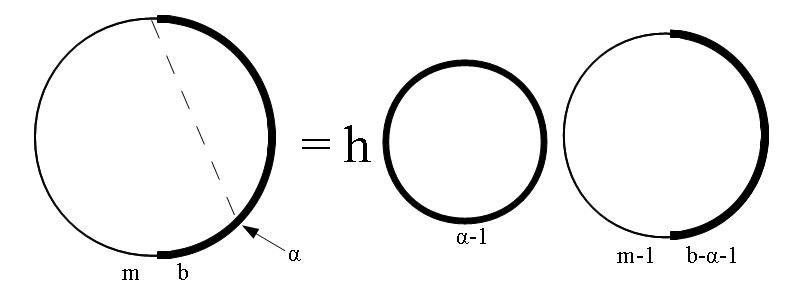}   
\end{center}
\vspace{-2mm}
\caption{Illustration of equation (\ref{12}). Circles represent expectation values of matrices. The thin line represents $M$'s, the solid line $B$'s and the dashed line represents the contraction. Planarity ensures that there are going to be no contractions between the two parts.}
\end{figure}  

We now turn to the correlator 
$\la \Tr ( M^m \, B^b ) \ra$. In evaluating this by Wick contractions, first consider 
the contraction of an $M$ matrix next to the series of $B$'s with one of the $B$'s.
The result may be written as (see Figure 1)
\be
\la  ( M^{m-1} )_{ij}\, {\dot M}_{jk} \, (B^\alpha)_{kr} \, {\dot B}_{rs} (B^{b-\alpha -1})_{si}\ra
= h~ {1\over N} \la \Tr B^\alpha \ra ~ \la \Tr ( M^{m-1} B^{b-\alpha-1})\ra
\label{12}
\ee
In arriving at this, we have used two results: Since nonplanar contractions are suppressed at large $N$, the contractions of the series of $B$'s in $B^\alpha$ have to be within themselves, and hence the series can be replaced by the expectation value. Further, since the expectation value is
proportional to the identity from (\ref{11}), the remaining set of matrices fall into a matrix product.
Finally, the contracted matrices, indicated by the overdots, give a factor of $h$.
Similarly, considering the contraction of the chosen $M$ matrix with another $M$ matrix, we get
\be
\la (M^\beta)_{ij}\, {\dot M}_{jk}\, (M^\alpha)_{kr} \,{\dot M}_{rs}\, (M^{m-2-\alpha-\beta} \, B^b)_{si}
\ra = f\,  {1\over N}\la  \Tr M^\alpha \ra\, \la \Tr (M^{m- 2 -\alpha} \, B^b)\ra
\label{13}
\ee
Combining (\ref{12}) and (\ref{13}) and allowing for all possible values of $\alpha$, we get the recursion rule
\bear
\la \Tr (M^m \, B^b) \ra &=& f\, {1\over N} \sum_{\alpha =0}^{m-2} \la \Tr M^\alpha \ra\, \la \Tr (M^{m-2-\alpha} B^b\ra)
\nonumber\\
&&
+\,  h \, {1\over N}\sum_{\alpha =0}^{b-1} \la \Tr (M^{m-1} \, B^{b-\alpha -1}) \ra\,
\la \Tr B^\alpha \ra
\label{14}
\eear
Since one $M$ is chosen to be contracted with all other matrices, this relation applies when $m > 0$.

In a similar way, we can single out a $B$ matrix adjacent to the series of $M$ matrices and consider its contractions. This leads to the recursion rule, for $b > 0$,
\bear
\la \Tr (M^m \, B^b )\ra &=& g \, {1\over N} \sum_{\alpha =0}^{b-2}
\la \Tr B^\alpha \ra \, \la \Tr ( M^m \, B^{b-2-\alpha} ) \ra\nonumber\\
&& 
+ \, h\, {1\over N}\sum_{\alpha =0}^{m-1} \la \Tr (M^{m-\alpha -1} \,B^{b-1})\ra\,
\la \Tr M^\alpha \ra
\label{15}
\eear
These recursion rules may also be viewed as the Schwinger-Dyson equations for
expectation values calculated via the functional integral
\be
\la {\cal O }\ra = \int [dM]\, e^{-S (M)}~ {\cal O}
\label{15a}
\ee
Equation (\ref{15}) can be obtained, for example, by considering the identity
\be
\int [dM]\, {\del \over \del M_{ji}} \left[ (M^{m-1}\, B^{b-1} )_{ji}\, e^{-S}\right] = 0 
\label{15b}
\ee
We will not consider the simplifications of such matrix integrals, but rather proceed to the direct solution of the recursion rules.
For this, we now define the normalized correlator
\be
W_{m,b} = \frac{1}{N}\left< \,  \Tr \left[ \left(\frac{M}{2\sqrt{f}}\right)^m \left(\frac{B}{2\sqrt{g}}\right)^b \right] \right>
\label{16}
\ee
The two recursion rules become
\be
4~W_{m,b} = \sum_{\alpha =0}^{m-2} W_{\alpha,0} W_{m-2-\alpha, b} 
+ \gamma \, \sum_{\alpha =0}^{b-1} W_{m-1, b-\alpha -1} W_{0,\alpha}
\label{17}
\ee
\be
4~W_{m,b} = \sum_{\alpha =0}^{b-2} W_{0,\alpha} W_{m, b-\alpha-2} 
+ \gamma \, \sum_{\alpha =0}^{m-1} W_{m- \alpha -1, b -1} W_{\alpha,0}
\label{18}
\ee
where
\be
\gamma = \frac{h}{ \sqrt{f\,g}}
\ee
Define the generating function
\be
\phi (t,s ) = \sum_{m, b =0}^\infty \, W_{m,b} \, t^m\, s^b
\label{19}
\ee
which is normalized as $\phi (0, 0) = W_{0,0} =1$.
The recursion rules now become
\bear
4 \, \left[ \phi (t, s) - \phi (0, s) \right] &=& t^2 \, \phi (t,s) \, \phi (t, 0) ~+~ \gamma \, t s \, \phi(t,s) \, \phi (0, s)
\nonumber\\
4\, \left[ \phi (t, s) - \phi (t, 0)\right] &=& s^2 \, \phi (t,s) \, \phi (0, s) ~+~ \gamma \, t s \, \phi(t,s) \, \phi (t, 0)
\label{20}
\eear
Solving these equations for $\phi (t,s )$ and equating the two expressions, we find
\be
4 \phi(t,0) - t^2 \phi(t,0)^2 = 4 \phi(0,s) - s^2 \phi(0,s)^2 
\label{21}
\ee
Thus each expression must be a constant, which should be $4$ from $\phi (0,0) =1$. We then solve
(\ref{21}) and use it in (\ref{20}) to get
\be
\phi(t,s) = \frac{4}{\left( 1+ \sqrt{1-t^2} \right) \left( 1+ \sqrt{1-s^2} \right) - \gamma ts}
\label{22}
\ee
Also, as a consequence of (\ref{7a}), we have for the constant $\gamma$:
\be
-1 \le \gamma \le 1
\label{22a}
\ee
The value of $\gamma$ is crucial. For $\gamma =0$, in particular,
the above generating function is the product of the generating
functions of two independent Wigner distributions, while for nonzero
$\gamma$ we have correlations.

\subsection{Scaling and special cases}

To explore the possible values of $\gamma$, it is useful to consider the scaling properties of the functions $f$, $g$, $h$.
Going back to their definition in (\ref{6}-\ref{8}), 
and taking $G(l)$ to go like $l^\alpha$ for large values of $l$, we see that
\bear
f &\sim& {1\over N} \sum (2 l +1) \, G(l) \sim {2\over N} \sum l^{\alpha +1}\nonumber\\
g&\sim& {2\over N} \sum l^{\alpha+5}, \hskip .2in h \sim {2\over N} \sum l^{\alpha +3}
\label{23}
\eear
The summations can be approximated by integrations when the exponent is larger than
$-1$; otherwise they are dominated by small values of $l$.
From the behavior in (\ref{23}), we see that there are four possible cases.
\begin{enumerate}
\item $\underline{\alpha > -2}$:\\
In this case $f\sim  N^{\alpha+1}$, $g\sim N^{\alpha + 5}$, $h\sim N^{\alpha +3}$
and so, $\gamma \sim 1$ (order $N^0$).
\item $\underline{ - 4 < \alpha < -2}$:\\
In this case $f\sim N^{-1}$, $g\sim N^{\alpha +5}$, $h \sim N^{\alpha +3}$ and so,
$\gamma \sim N^{{\half} (\alpha + 2)}  \ll 1$.
\item $\underline{-6 < \alpha < -4}$:\\
This leads to $f \sim N^{-1}$, $g \sim N^{\alpha +5}$, $h \sim N^{-1}$, and hence,
$\gamma \sim N^{- {\half} ( \alpha + 6)} \ll 1$.
\item $\underline{a < -6}$:\\
In this case, $f \sim N^{-1}$, $g \sim N^{-1}$, $h\sim N^{-1}$ and hence,
$\gamma \sim 1$.
\end{enumerate}
In cases 2 and 3, the mixed term will be irrelevant  ($\gamma \rightarrow 0$) and $\phi (t,s)$ becomes the product of the moment generating functions for two independent Wigner distributions.
In the other two cases, $\gamma \sim 1$, and the distribution will be different from the Wigner semicircle law.
The case of $\alpha = - 2$ is special. In this case, $\gamma$ vanishes as
$(\ln N )^{-1}$. The distribution function will have to be evaluated by taking a
proper limit as $\gamma$ tends to this value. Notice that $\alpha  =-2$ is the case for the
kinetic term being the Laplacian. A similar statement applies to $\alpha = - 6$.

\subsection{The distribution function}

In oder to obtain the distribution function from the moments, we start
with (\ref{22}) and expand $\phi (t,s )$ in $\gamma$ as
\be
\phi(t,s) = \sum_{n=0}^\infty \left(\frac{\gamma}{4}\right)^n t^n \phi(t)^{n+1} s^n \phi(s)^{n+1}
\label{24}
\ee
where
\be
\phi(u) = \phi(u,0) = \phi(0,u) = \frac{2}{ 1+ \sqrt{1-u^2}} = 2\frac{1-\sqrt{1-u^2}}{u^2}
\label{25}
\ee
So to each order in $\gamma$ the distribution factorizes as
\be
\rho(x,y) = \sum_{n=0}^\infty \left(\frac{\gamma}{4}\right)^n\, \rho_n (x)\, \rho_n (y)
\label{fac}
\ee
To find the $n$-th order distribution $\rho_n (x)$ corresponding to the generating function $t^n \phi(t)^{n+1}$ we work as follows:
Define $[f]_+$ the non-negative power (non-singular) part of a function
\be
\left[ \dots + a_{-2} t^{-2} + a_{-1} t^{-1} +  a_0 + a_1 t + a_2 t^2 + \dots \right]_+
= a_0 + a_1 t + a_2 t^2 + \dots
\label{27}
\ee
Then, if the generating function $f(t)$ corresponds to the distribution $w(x)$, the generating function
\be
\left[ \frac{f(t)}{t^n} \right]_+ ~, ~~n\ge 0 ~~~\to ~~~ x^n w(x)
\label{sing}
\ee
corresponds to the distribution $x^n w(x)$, as can easily be shown.
(Note that this is not true for $n<0$.)

Now, it is easily verified by direct substitution that the function
$\phi (t)$ obeys the identity
\bear
\phi(t)^{n+1} &=& \left(\frac{2}{t^2}\right)^n\Biggl[ \frac{1}{2\sqrt{1-t^2}} \left[ \left( 1 + \sqrt{1-t^2} \right)^{n+1}
- \left( 1 - \sqrt{1-t^2} \right)^{n+1} \right] \phi(t) \cr
&&\hskip .5in - \frac{1}{\sqrt{1-t^2}} \left[ \left( 1 + \sqrt{1-t^2} \right)^n
- \left( 1 - \sqrt{1-t^2} \right)^n \right] \Biggr]
\label{id}
\eear
Upon expanding the binomials and
multiplying with $t^n$ we obtain
\be
t^n \phi(t)^{n+1} = \sum_{k=0}^{[n/2]} \left( \begin{array}{cc} n+1 \cr 2k+1 \end{array} \right) \left(\frac{2}{t}\right)^n
(1-t^2)^k \, \phi(t)
- 2  \sum_{k=0}^{[(n-1)/2]} \left( \begin{array}{cc} n \cr 2k+1 \end{array} \right) \left(\frac{2}{t}\right)^n
(1-t^2)^k
\label{28}
\ee
We observe that the coefficient of $\phi(t)$ contains only negative or zero powers of $t$, while the second term
contains only negative powers. The left hand side, however, is obviously nonsingular. Therefore, the second term
cancels the singular part of the first term but does not contribute to the nonsingular part. 
Using the result (\ref{sing}) we obtain
\be
\rho_n (x) = \sum_{k=0}^{[n/2]} \left( \begin{array}{cc} n+1 \cr 2k+1 \end{array} \right) 2^n x^{n-2k}
(x^2-1)^k \, \rho(x)
\label{rn}
\ee
where $\rho (x)$ is the Wigner distribution,
\be
\rho(x) = \frac{2}{\pi} \sqrt{1-x^2}
\label{29}
\ee

We can now consider the two-dimensional distribution 
$\rho(x,y)$ as defined by
\be
W_{mb} = \int dx \,dy\, \rho(x,y) ~ x^m \,y^b  \label{30}
\ee
The distribution $\rho(x, y)$ can be calculated by substituting for
$\rho_n $ from (\ref{rn}) in (\ref{fac}) and carrying out the summation.
This task is facilitated by noticing that
(\ref{rn}) can be obtained from the first term in (\ref{id}) by substituting $t=x^{-1}$. After some algebra we find
\be
\rho(x,y) = \rho(x) \rho(y) \frac{1-\gamma^2}{(1-\gamma^2)^2 - 4\gamma(1+\gamma^2)xy + 4\gamma^2(x^2 + y^2)}
\label{distr}
\ee
This is symmetric in $x$ and $y$. The integral of $\rho(x,y)$ over $y$ is nontrivial; it can be calculated and gives $\rho(x)$ as it should, giving a check on the expression for
$\rho (x, y)$.

Although the quadratic form in $x,y$ in the denominator is not positive definite, the above distribution is positive
for $\gamma^2 \le 1$. Indeed, for $\gamma >0$ the negative eigenvalue of the quadratic form corresponds to
the eigenvector $x=y$. Putting $x=y=1$, their maximal value given the Wigner distribution prefactor $\rho(x) \rho(y)$,
the denominator becomes $(1-\gamma)^4$. Similarly, for $\gamma <0$ the negative eigenvalue corresponds to
$x=-y$ and choosing $x=-y=1$ the denominator becomes $(1+\gamma)^4$. For $\gamma^2=1$ the distribution
appears to be singular, but taking the limit we see that it becomes $\rho(x) \delta (x-y)$ for positive $\gamma$ and
$\rho(x) \delta (x+y)$ for negative $\gamma$.

For $\gamma^2 >1$, however, the distribution becomes negative, signaling the nonexistence of a probability
interpretation in that case. We can understand this by noticing that the correlation of the variables $x$ and $y$
is calculated as
\be
\frac{\left< xy \right>}{\sqrt{\left< x^2 \right> \left< y^2 \right>}} = 
\left. \frac{2\partial_t \partial_s \phi}{\sqrt{\partial_t^2 \phi \, \partial_s^2 \phi}} \right|_{t=s=0} = \gamma
\label{31}
\ee
Since correlations of stochastic variables have to be between $-1$ and $1$, this is also the allowed range of $\gamma$.
For $\gamma=\pm 1$ the two variables are fully (anti)correlated and thus equal to (minus) each other, justifying
the delta-functions.

The relation (\ref{22a}) ensures that the distribution is always positive. For instance, since $\gamma$ is the ratio $h/\sqrt{fg}$,
in the regular large-$N$ scaling case 
corresponding to $G(\ell ) \sim N^\alpha$, $\alpha >-2$ (case 1 of subsection 2.3),  it becomes
\be
\gamma = \frac{\sqrt{(\alpha+2)(\alpha+6)}}{\alpha+4}
\label{32}
\ee
We see that $\gamma$ is always between $0$ and $1$, and it tends to $1$ in the limit $\alpha \to \infty$. In this limit, it is 
reasonable that the eigenvalues of $M$ and $B$ are correlated since they are dominated by the
largest angular momentum sector.

In conclusion, we obtain a correlated distribution for the eigenvalues of $M$ and $B$, while their
marginal distributions remain Wigner semicircles of radii $2\sqrt f$ for $M$ and $2\sqrt g$ for $B$,
as seen from the scaling factors in (\ref{16}).

\section{The massless case}

The action describing a massless scalar on the fuzzy sphere, $\mu=0$ in
the action (\ref{1}), merits special attention. This is the critical
case $\alpha = -2$ in the scaling of $G(l)$ at large $l$, corresponding
to a very weak vanishing of $\gamma \sim (\ln N)^{-1/2}$ at the large $N$ limit.
 Further, for $\mu =0$ the constant mode of the field (the trace of $M$) 
drops out of the action and must be eliminated from the calculation.

In fact, taking the limit $\mu \to 0$ in this case is somewhat nontrivial.
The propagator is $G(l) = [\mu^2 + l (l+1)]^{-1}$, and for 
$\mu \sim 1/\ln N$ or smaller we obtain in the large $N$ limit
\be
f = \frac{1}{N} (\mu^{-1} + 2\ln N )~,~~ g = \half N^3 ~,~~ h = N
\ee
which leads to a Wigner radius of $2 {\sqrt f} = 2
\sqrt{(\mu^{-1} + 2\ln N )/N}$
for the eigenvalue distribution of $M$. This not only diverges
as $\mu$ goes to zero, but is also misleading: for such low
values of $\mu$ the planarity property of matrix expectation values
fails, since the trace part of $M$ contributes to the same order
(or higher) than the traceless part and arises in all diagrams
(planar and nonplanar). Such contributions give rise to a Gaussian,
rather than Wigner, distribution.

To understand this better, we decompose $M$ in its trace part $c_0$
and its traceless ($l \neq 0$) part $\tilde M$
\be
M = \frac{c_0}{\sqrt N} + {\tilde M} = \frac{c_0}{\sqrt N}
+ \sum_{l>0,\, A}c_A^l \, T^{(l)}_A
\ee
The trace and traceless parts decouple. The eigenvalues of $\tilde M$
have a Wigner distribution at large $N$ with radius 
$2 \sqrt{2\ln N /N}$,
arising from $f$ with $l=0$ dropped, while the single mode of the
trace part contributes a shift distributed as a Gaussian with spread
$1/\sqrt{N\mu}$. Since the two distributions are independent, the
total eigenvalue distribution will be given by their convolution.
For $\mu \gg 1/\ln N$ the spread of the Gaussian is much smaller than
the Wigner radius and the convolution essentially gives back the Wigner.
For $\mu \ll 1/\ln N$, on the other hand, the Gaussian dominates.
For $\mu \sim 1/\ln N$ we get an intermediate distribution.

In order to obtain the massless result, therefore, we have to omit the
trace part of the matrix (which has vanishing action) and substitute
$\tilde M$ for $M$. The result is again a Wigner semicircle of radius
$\sqrt{8\ln N /N}$  for $\tilde M$, correlated weakly ($\gamma \sim (\ln N)^{-1/2}$)
with a Wigner semicircle for $B$ of radius $2\sqrt{g} = \sqrt{2 N^3}$.
To first order in $\gamma$, odd-odd expectation values of the form $W_{2k+1,2a+1}$
also survive and are given by $(\gamma/4) \rho_1 (x) \, \rho_1 (y)$
as in (\ref{fac}).

\section{Conclusions and discussion}

We recovered a correlated distribution that reproduces the expectation values of (ordered) matrix products of the
basic matrix variable $M$ and its Laplacian $B$ in the large-$N$ limit. This is our main result.

The distribution of eigenvalues of $M$ remains a Wigner semicircle, but with a renormalized radius $R = 2 \sqrt f$,
as is evident from (\ref{16}) and (\ref{distr}). Similarly, the distribution of eigenvalues of $B$ is also a Wigner
semicircle with radius $R' = 2 \sqrt g$. In this sense, the Wigner semicircle is very robust, arising basically from
the planarity property of expectation values of matrix observables in the large-$N$ limit. This result is at odds
with the result of \cite{ocon} which gives a polynomially deformed Wigner distribution in the presence of
the kinetic term in the action. The source of the discrepancy could be the fact that that the kinetic term becomes 
dominant in the large-$N$ limit, while it was treated perturbatively in \cite{ocon}.

The new element in our result is the correlation $\gamma$. The distribution (\ref{distr}) is a two-dimensional
correlated Wigner distribution (whose one-dimensional marginals are the standard Wigner semicircle). Indeed,
the expectation values of mixed products of $M$ and $B$  do depend on the angular variables of $M$ and
these introduce the nontrivial (and non-unit) correlations. This is independent of the exact form of the action,
as long as the trace sector does not dominate and $\gamma$ remains nonzero in the large-$N$ limit.
 For instance, for the standard matrix model
(without the kinetic term), that is, the case $\alpha = 0$ in the scaling of $G(l)$, we have
$\gamma = \sqrt{3}/2$. For the case with the kinetic term, $\alpha = -2$, the correlation vanishes weakly
(logarithmically) in the large-$N$ limit.

There are obviously many issues remaining to be analyzed; namely the expectation value of more general
products of matrix variables, the question of the dependence of the results on the ordering of matrices and
the relevance of the results to string, membrane or gravity models. These will be examined in forthcoming
publications.

Further, we point out that our distribution represents an explicit expression of {\it correlated} free variables in
Voiculescu's noncommutative probability theory. Large-$N$ matrix models with general propagator $G(l)$,
and their generalized matrix observables, are a natural arena where these mathematical notions are
realized and make their physical significance clear.

The direct physical relevance of our results lies mainly in
noncommutative theory and its properties, contrasted to those
of a regularized standard (commutative) field theory. We see
that the noncommutative theory presents qualitatively different
features, mainly related to planarity, which is itself an expression
of UV/IR mixing. The generic emergence of Wigner, rather than gaussian,
distributions, and their correlated generalizations, 
is the key signature
of this effect. Other possible physical applications of our results
range from quantum Hall situations to quantum gravity. Matrix models
for the quantum Hall effect have been proposed and used to some
advantage \cite{QH}, 
and the inclusion of a kinetic term to such models adds
an element of compressibility that could be probed by an approch
similar to the one in this work. Finally, quantum gravity remains the
main motivation behind this work and the possibility to probe 
Planck-scale
effects, such as the formation and evaporation of microscopic black
holes, without the benefit (or burden) of string theory is an
exciting prospect.

\vskip 0.1in

{\it \underline {Acknowledgements}:} This research was supported by National Science Foundation grant PHY-0855515 and by PSC-CUNY grants.

\end{document}